\documentclass[12pt,a4paper]{article}
\usepackage{amsmath}
 \usepackage{bm}
 \usepackage{natbib}
  \usepackage{setspace}
  \usepackage{color}
   \usepackage{graphicx}
  \usepackage{subfigure}
  \usepackage{algorithm}
  \usepackage{algpseudocode}
  \usepackage{multirow}
  \usepackage{mathtools}  
  \usepackage{url}
  \usepackage{authblk}
  \DeclarePairedDelimiter{\ceil}{\lceil}{\rceil}
  \algblock{ParFor}{EndParFor}
  \algnewcommand\algorithmicparfor{\textbf{parfor}}
  \algnewcommand\algorithmicpardo{\textbf{do}}
  \algnewcommand\algorithmicendparfor{\textbf{end\ parfor}}
  \algrenewtext{ParFor}[1]{\algorithmicparfor\ #1\ \algorithmicpardo}
  \algrenewtext{EndParFor}{\algorithmicendparfor}
  \newcommand{\longoverbrace}[2]{\overbrace{#1}^{\text{\hbox to 0cm{\hss #2 \hss}}}}  
  \newcommand{\longunderbrace}[2]{\underbrace{#1}_{\text{\hbox to 0cm{\hss #2 \hss}}}}
  \newcommand{\xb}{\bm{x}}
  \newcommand{\yb}{\bm{y}}  
  \newcommand{\wb}{\bm{w}}
  \newcommand{\ub}{\bm{u}}  
  \newcommand{\nb}{\bm{n}}  
  \newcommand{\xib}{\bm{\xi}} 
  \newcommand{\betab}{\bm{\beta}}
  \newcommand{\thetab}{\bm{\theta}}
  
  \newcommand{\ths}{\theta^\star}
  \newcommand{\thsb}{\bm{\theta^\star}}
  \newcommand{\mub}{\bm{\mu}}
  \newcommand{\Sigmab}{\bm{\Sigma}}
  \renewcommand{\th}{\theta}
\newcommand{\yt}{\tilde{y}}
\newcommand{\ytb}{\bm{\tilde{y}}}
\newcommand{\nt}{n}
\newcommand{\ntb}{\bm{n}}
\newcommand{\rt}{r}
\newcommand{\rtb}{\bm{r}}
\newcommand{\st}{\tilde{s}}

\newcommand{\stib}{\bm{\tilde{s}^{-i}}}

\newcommand{\thts}{\theta^\star}
\newcommand{\thtsb}{\bm{\theta^\star}}
\newcommand{\St}{S}
\newcommand{\rhot}{\rho}

\title{\bf Scalable Bayesian Nonparametric Clustering and Classification}

\author[1]{Yang Ni}
\author[2]{Peter M\"uller}
\author[1]{Maurice Diesendruck}
\author[3]{Sinead Williamson}
\author[4]{Yitan Zhu}
\author[4,5]{Yuan Ji}
\affil[1]{Department of Statistics and Data Sciences, The University of Texas at Austin}
\affil[2]{Department of Mathematics, The University of Texas at Austin}
\affil[3]{Department of Information, Risk, and Operations Management, The University of Texas at Austin}
\affil[4]{Program for Computational Genomics and Medicine, NorthShore University HealthSystem}	
\affil[5]{Department of Public Health Sciences, The University of Chicago}

\date{}

\begin{document}
\def\spacingset#1{\renewcommand{\baselinestretch}%
	{#1}\small\normalsize} \spacingset{1}
\maketitle

\bigskip

\begin{abstract}
We develop a scalable multi-step Monte Carlo algorithm for inference under a large class of nonparametric Bayesian models for clustering and classification. Each step is ``embarrassingly parallel" and can be implemented using the same Markov chain Monte Carlo sampler. The simplicity and generality of our approach makes inference for a wide range of Bayesian nonparametric mixture models applicable to large datasets. Specifically, we apply the approach to inference under a product partition model with regression on covariates. We show results for inference with two motivating data sets: a large set of electronic health records (EHR) and a bank telemarketing dataset. We find interesting clusters and favorable classification performance relative to other widely used competing classifiers.
%The proposed algorithm is simulation exact in the sense that the inference is carried out under the posterior model that is based on the entire dataset.
\end{abstract}

\noindent%
\vfill

\newpage
\spacingset{1.45} % DON'T change the spacing!
\section{Introduction}
We propose a distributed Monte Carlo algorithm for 
Bayesian nonparametric clustering and classification methods that are
suitable for data with large sample size. The algorithm is applicable
for both conjugate and non-conjugate structures, and consists of
$K$ computationally efficient steps.   $K$ is dynamically determined and is typically less than 4.   In each
of the first $(K-1)$ steps, we divide the data into many shards and run
``embarrassingly parallel" Markov chain Monte Carlo (MCMC) simulations
in each shard. In the last step, MCMC is run again to generate approximate samples from the full posterior. We apply the
algorithm for inference in a
product partition model with regression on covariates (PPMx, 
\citealt{muller2011product}), and show results for a large
electronic health records (EHR) dataset and a dataset of telemarketing for long-term bank deposits. Our method is scalable, outperforms
state-of-the-art classifiers and generates interpretable partitions of
the data.

\paragraph*{Classification and clustering.}
We consider Bayesian nonparametric (BNP) methods for clustering
and classification. Classification aims to assign observations into two or more categories on the basis of training data
with known categories.  Widely used
classification algorithms include logistic regression (LR), naive
Bayes, neural networks, k-nearest neighbors, support vector machines
(SVM, \citealt{cortes1995support}), decision trees, random
forests (RF, \citealt{ho1995random}), classification and regression
trees (\citealt{breiman1984classification}), Bayesian additive
regression trees (BART, \citealt{chipman2010bart}) and   mixture
models based on Bayesian nonparametric (BNP) priors. Some recent examples for the latter are
\cite{cruz2007semiparametric} who use a dependent Dirichlet process prior,
\cite{mansinghka2007aclass} who model the distribution within each
subpopulation defined by the class labels using a Dirichlet process
mixture model, or  
\cite{gutierrez2014bayesian} who use a geometric-weights prior
instead. For more examples, see a recent review by \cite{singh2016review} and
references therein. 

In contrast to supervised learning in classification, 
clustering methods partition the observations into latent groups/clusters in an
unsupervised manner, with the aim of creating homogeneous
groups such that observations in the same cluster are more
similar to each other than to those in other clusters. Widely used clustering methods include hierarchical clustering, k-means, DBSCAN
\citep{ester1996density} and finite mixture models. Posterior simulation for finite mixtures was first discussed in \cite{richardson1997bayesian} and extended to multivariate
mixtures in \cite{dellaportas2006multivariate}.
% ,nobile2004posterior,tadesse2005bayesian,
% dellaportas2006multivariate,miller2017mixture}.
See, for example, \cite{jain2010data} and \cite{fahad2014survey} for
recent reviews. 

BNP \citep{hjort2010bayesian} clustering methods 
offer a wide range of flexible alternatives to classical clustering
algorithms including Dirichlet process mixtures (DPM,
\citep{lo1984class,maceachern2000dependent,lau2007bayesian} and variations with different data structures, such as
\citep{rodriguez2011sparse} for a mixture of graphical models,
% \citep{rodriguez2008nested}
% dahl2006model,,,ICML2012Kulis_291,
Pitman-Yor (PY) process mixtures
\citep{pitman1997two, %ishwaran2001gibbs, 
ni2016heterogeneous},
normalized inverse Gaussian process mixtures
\citep{lijoi2005hierarchical}, normalized generalized Gamma process
mixtures \citep{lijoi2007controlling}, and more general classes of BNP mixture models
\citep{barrios&al:13,favaro&teh:13,Argiento:2010}. 
 
\paragraph{Scalable methods.}
Datasets that are too large to be analyzed on a single machine
increasingly occur in many applications, 
including health care, online streaming, social
media, education, banking and finance. Many of the
earlier mentioned classification or clustering methods do not scale to
large datasets, partly due to lack of straightforward parallelization. Below, we 
briefly review some recently proposed 
efficient computational strategies.

 \cite{zhang2012communication}
developed two algorithms for parallel statistical optimization based
on averaging and bootstrapping. \cite{kleiner2014scalable} developed a
scalable bootstrap to evaluate the uncertainty of estimators.

 Bayesian methods naturally provide uncertainty quantification of estimators but
are in general computation-intensive. \cite{huang2005sampling}
proposed consensus Monte Carlo algorithms that distribute data to
multiple machines running separate MCMC simulations in parallel. 
Various ways of eventually consolidating simulations from these
subset posteriors have been proposed
\citep{neiswanger2013asymptotically,wang2013parallelizing,white2015piecewise,minsker2014scalable,scott2016bayes}. 
An alternative strategy for scalable Bayesian computation is
based on approximating the full likelihood/posterior using subsampling
techniques
\citep{welling2011bayesian,korattikara2014austerity,bardenet2014towards,quiroz2018speeding};
see \cite{bardenet2015markov} for a review of related recent MCMC
approaches. Alternatively to MCMC, Bayesian inference can be 
carried out by using approximation such as variational Bayes
\citep{jaakkola2000bayesian,ghahramani2001propagation,broderick2013streaming,hoffman2013stochastic}. For a grand overview of Bayesian computation, see also \cite{green2015bayesian}.
Although variational inference is scalable to large-scale
datasets and usually yields good approximations to the marginal
posterior, 
MCMC algorithms tend to better approximate the joint posterior because they are simulation-exact methods.

\paragraph{Scalable classification and clustering.}
Some classical classifiers like logistic regression are scalable to
large datasets and easy to interpret. However, the performance of
logistic regression tends to be not as accurate as other ``black box"
classifiers. Ideally, a good classifier does not need to sacrifice its
predictive performance for interpretability and scalability. This is
what we aim to achieve in this paper.

Some work has been done in this area. \cite{payne2014bayesian} developed a
two-stage Metropolis-Hastings algorithm for logistic regression to avoid the need for exact likelihood
computation. The first stage, based on an approximate likelihood, is
used to determine whether a full likelihood evaluation is necessary in
the second stage. Combined with consensus Monte Carlo, the proposed
method scales well to datasets with large
samples. \cite{rebentrost2014quantum} 
implemented SVM on a
quantum computer and showed an exponential speed-up compared to
classical sampling algorithms.

%\cite{maceachern1999sequential} proposed a sequentialimportance sampling approach to implement posterior inference for random partitions,  without recourse to MCMC. 
For clustering, \cite{pennell2007fitting} developed a two-stage approach for
fitting random effects models to longitudinal data with large sample
size. They first cluster subjects using a deterministic algorithm
and then cluster the group-specific random effects using a DPM model. \cite{zhao2009parallel} proposed a parallel
k-means clustering algorithm using the MapReduce framework
\citep{dean2008mapreduce}. \cite{wang2011fast} developed a single-pass
sequential algorithm for conjugate DPM models. In each iteration, they
deterministically assign new subject to the cluster with the highest
probability conditional on past cluster assignments and the data up to
current observation. The algorithm is repeated for multiple permutations of the
samples. \cite{Lin:13} also proposed a one-pass sequential algorithm
for DPM models. The algorithm utilizes a constructive characterization of
the posterior distribution of the mixing distribution given data and
partition.
% Marginalizing out the random partition, the desired
% posterior distribution is obtained. 
Variational inference is adopted
to sequentially approximate the
marginalization. \cite{WilliamsonAl:13} introduced a parallel MCMC for
DPM models which involves iteration over local updates and a global update. For
the local update, they exploit the fact that Dirichlet mixtures of
Dirichlet process (DP) are DP if the parameters of Dirichlet mixture are suitably chosen. 
% Based on this fact, they
% augmented the model with auxiliary variables which distribute samples
% to different processors. Conditioned on these auxiliary variables, the
% samples from one processor are conditional independent of all other
% samples, which allows for parallelizable local inference. Then a
% global update is performed by sampling the auxiliary variables to move
% data across processors. 
\cite{GeAl:15} used a similar characterization
of the DP as in \cite{Lin:13}. But instead of variational approximation,
they adapted the slice sampler for parallel computing under a
MapReduce framework. \cite{TankAl:15} developed two variational
inference algorithms for general BNP mixture models.
% The first algorithm is a
% single-pass algorithm for streaming/sequential data based on assumed
% density filtering (ADF). ADF allows for efficient sequential updates
% while keeping the infinite-dimensional nature of the BNP model. Then
% they extended it to a multi-pass algorithm for fixed-size data by
% leveraging the connection between  ADF and expectation
% propagation.

  The method most similar to that proposed in this paper is the subset
  nonparametric Bayesian (SNOB) clustering of
  \cite{zuanetti2018Bayesian}, a computation-efficient alternative 
for model-based clustering under a DPM model with
conjugate priors.  
% The first approach is based on a
% predictive recursion algorithm \citep{newton1998nonparametric} which
% only a single scan of all observations and avoids expensive
% MCMCs. 
SNOB is a two-step approach. It first splits data into shards and computes the clusters
locally in parallel. A second step combines the local
clusters into global clusters. All steps are carried out using MCMC simulation under a common
DPM model. However, the method requires conjugate models.
%  using regular MCMC. Since continuous
% cluster-specific parameters are almost surely non-identical, all
% cluster-specific parameters are marginalized out in the second
% step to allow merging two local clusters. The latter requires conjugate models and limits its
% applicability in a wide range of applications where non-conjugate models are desired. 

\paragraph{Proposed method.}
Inspired by Neal's algorithm 8 \citep{neal2000markov}   for inference in DPM models,   
we extend SNOB to clustering under non-conjugate BNP models, and propose a multi-step
algorithm for \underline{s}ubset
\underline{i}nference of \underline{g}eneral \underline{n}onparametric
Bayesian methods (SIGN). The algorithm is a $K$-step approach   ($K$ is dynamically determined and will be introduced in Section \ref{sec:sign})  .
Each step requires computationally intensive clustering on small subsets
only. The number of required subsets is linear in the sample size $n$,
making it possible to implement posterior inference also for 
data that is too large to allow the use of full MCMC simulation. SIGN
can be applied with a large class of BNP  mixture models. Particularly, we show
how SIGN is implemented for inference under the PPMx model to
simultaneously cluster and classify patients from a large Chinese EHR
dataset with 85,021 samples and customers from a bank telemarketing
dataset with 37,078 records. 

In  the context of a classification problem, SIGN still
requires that all data can be accessed. This is not an inherent constraint of the proposed algorithm; rather it
is due to the lack of
sufficient/summary statistics for classification models (such as
probit regression). Whenever such statistics exist, SIGN does not need
to access the entire dataset. 

The remainder of this paper is organized as follows.  In Section
\ref{sec:sign}, we introduce the proposed SIGN algorithm which is applied for inference under the
PPMx model in Section \ref{sec:ppmx}. 
The  SIGN
algorithm is evaluated with simulation studies in Section
\ref{sec:cs} and applied to EHR and bank telemarketing data in Section
\ref{sec:cs2}. We conclude 
with a discussion in Section \ref{sec:disc}. 

\section{The proposed SIGN algorithm}
\label{sec:sign}
\subsection{BNP clustering}

We propose an algorithm for posterior inference on random partitions
under BNP mixture models. To state the general model, we need some notation.
A partition $\rho=\{S_1,\dots,S_C\}$ of an index
set $[n]=\{1,\dots,n\}$ is a collection of nonempty, disjoint and
exhaustive subsets $S_c \subseteq [n]$. The partition can
alternatively be represented by a set of cluster membership indicators
% characterized by the cluster memberships 
$\bm{s}=(s_1,\dots,s_n)$ with $s_i=c$ if $i\in S_c$.
Throughout the paper, we will use superscript $-i$ to represent
the appropriate quantity with the $i$th sample or the $i$th item
(defined later) removed. For instance,
$\bm{s}^{-i}=\bm{s}\backslash s_i$ and $\rho^{-i}=(\rho\backslash
S_{s_i})\cup (S_{s_i}\backslash i)$ are the cluster memberships and
partition after removing index $i$.

 In what follows we consider a \textit{random} partition $\rho$ with prior probability distribution
$p(\rho)$.
Let $n_c=|S_c|$ denote the cardinalities of
the partitioning subsets. Let $\nb=(n_1,\ldots,n_C)$  and let $\nb^{j+}$ denote $\nb$ with the $j$th element incremented by 1. The random partition is called exchangeable if 
$p(\rho)=f(\nb)$ for a symmetric (in its arguments) function
$f(\nb)$ and if $f(\nb)=\sum_{j=1}^{C+1} f(\nb^{j+})$. The function $f(\nb)$ is known as the
exchangeable partition probability function (EPPF). By Kingman's representation theorem \citep{Kingman:78}, any exchangeable
random partition can be  characterized as the groups formed by ties under
i.i.d. sampling from a discrete probability measure $G=\sum_{h=1}^\infty w_h \delta_{m_h}$.
That is, $\rho$ is determined by the ties among
$\th_i \sim G$, $i=1,\ldots,n$. 
We denote the unique values of $\theta_i$'s by
$\theta_1^\star,\dots,\theta_C^\star$, implying 
$i \in S_c$ if $\theta_i=\theta_c^\star$.  
See, for example, \cite{LeeAl:13} for a discussion.
It follows that a prior probability model for an exchangeable random partition $\rho$ can always be defined as a prior $p(G)$ on
a random discrete distribution $G=\sum_{h=1}^\infty w_h
\delta_{m_h}$. This implicit definition of $p(\rho)$
by a BNP prior $p(G)$ on the random probability measure $G$ is 
a commonly used specification of random partition models. The
construction already includes cluster-specific parameters $\ths_c$
which are useful for the construction of a sampling model conditional on
the partition. We use it in the next step of the model construction.

The model on $G$ and $\th_i$ is completed with a sampling model
for the observed data conditional on $\rho$.
For example, the $\th_i$ could index a sampling model $p(y_i \mid
\th_i)$, implying that all observations in a cluster share the same
sampling model.  In summary,
% A broad class of BNP models can be set up as a mixture model,
% e.g. Gibbs-type priors \citep{de2015gibbs}, 
\begin{eqnarray*}
\label{eqn:mixt}
y_i|\theta_i\sim p(y_i|\theta_i),\mbox{~~~~}\theta_i|G\sim G,\mbox{~~~~} G\sim H,
\end{eqnarray*}
where
$G$ is a discrete prior distribution of $\theta_i$ and $H$ is the
BNP prior for the random probability measure $G$. 

  There are a number of options for $H$. A popular choice is the DP, which yields an EPPF of the form $p(\rho|\nb)\propto\alpha^{C-1}\prod_{c=1}^C(n_c-1)!$ where $\alpha$ is the concentration parameter. Other choices include PY, the normalized inverse Gaussian process and the normalized 
generalized gamma process. In many applications, the focus is on the
posterior distribution of the random partition $\rho$, which can be
approximated by various MCMC algorithms, including, among many others, 
\cite{escobar1994estimating}, \cite{maceachern1998estimating}, 
\cite{neal2000markov} and \cite{walker2007sampling}, and -- for more
general models -- 
\cite{barrios&al:13} or \cite{favaro&teh:13}. 
However, MCMC is only practicable for small to moderate datasets. Directly applying MCMC to large
datasets is very costly because the algorithm has to scan through all
the observations at every iteration, each requiring likelihood and
prior evaluations. We consider inference on $\rho$ with large sample
size in next section.  

\subsection{SIGN algorithm}
\label{sec:sign}
\begin{figure}[h]
	\centering
	\includegraphics[width=.8\textwidth]{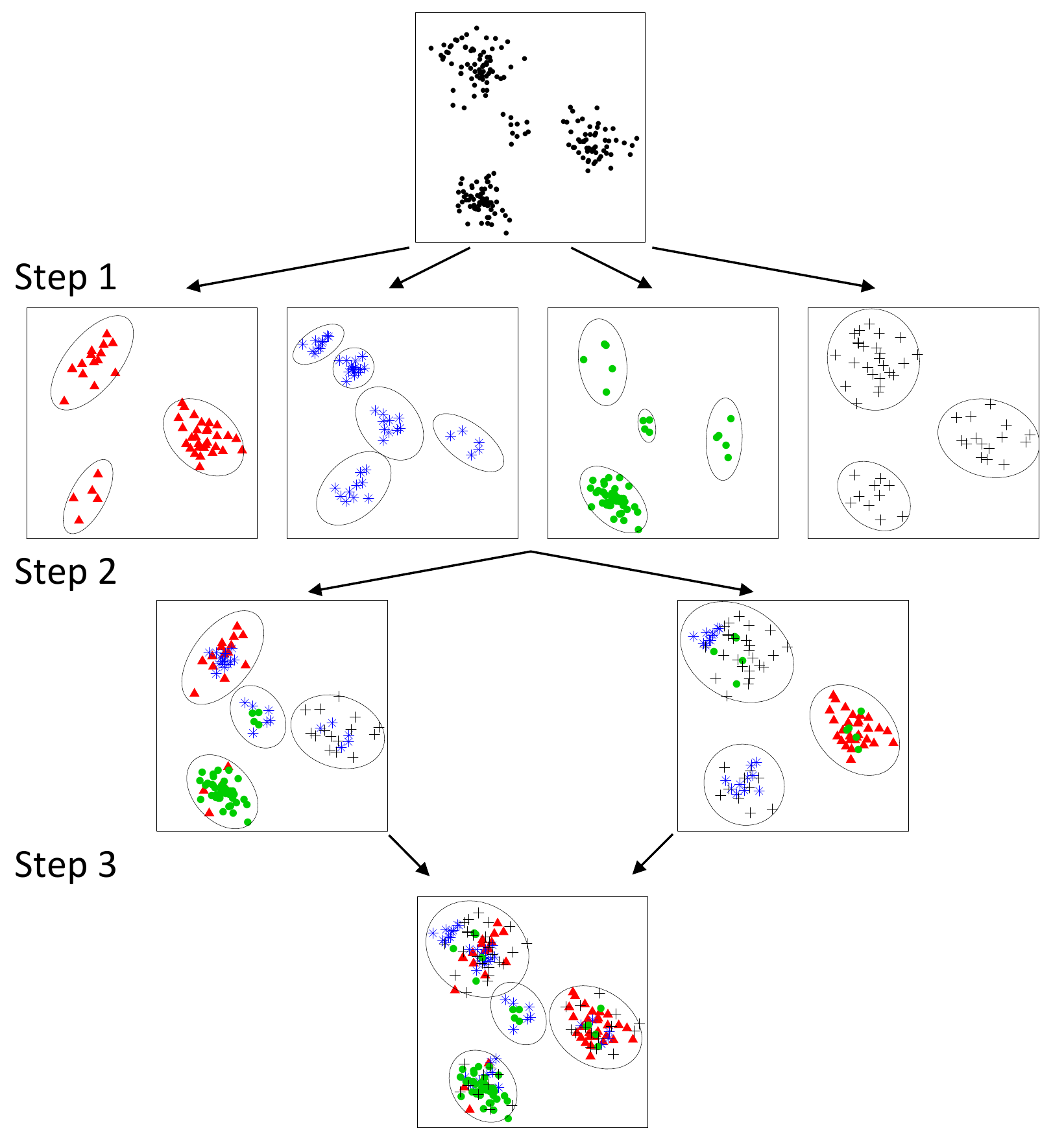}
	\caption{Example workflow of a 3-step SIGN algorithm. Step 1:
		the dataset is randomly distributed into 4 shards, each denoted by a
		unique type (color) of marker and observations are partitioned into
		local clusters (represented by the ellipses) within each shard in
		parallel. Step 2: local clusters are randomly distributed into 2
		shards and partitioned into regional clusters within each shard. Step
		3: regional clusters are aggregated and partitioned into global
		clusters.}
	\label{ill}
\end{figure}

The proposed SIGN algorithm proceeds in steps.  For illustration, an
example workflow of SIGN with $K=3$ steps is shown in Figure
\ref{ill}. Importantly, across all steps of the algorithm, all updates of
cluster configurations (initially of observations, and of sets of
observations in later steps) are based on a single underlying BNP
mixture model for the data. Details of the implied probabilities for
clustering sets of obserations are given later.  

\underline{Step 1.}
In the first step, the full dataset is randomly split into $M_1=4$
shards; the observations from each shard are denoted by a distinct
symbol in the figure. A clustering algorithm  (Neal's algorithm 8, in
our case)   is then applied to 
cluster the items (initially, in the first step, the observations) in
each shard separately, and can be implemented in parallel. We refer to the 
estimated clusters, represented by the ellipses, as ``local"
clusters. These local clusters are frozen, meaning that the
observations within each cluster will never be split in the subsequent
steps although merging is possible.

\underline{Step 2.}
In the second step, the local clusters estimated from the previous step become the
items to be clustered in the next step. That is, we freeze the earlier clusters, and only allow earlier
clusters to combine to larger clusters. 
The items are again split into $M_2$ shards ($M_2=2$ in Figure 1), and
are again clustered within each shard, still using the same underlying BNP mixture model. See later for
a statement of the relevant probabilities to cluster lower level
clusters. At the end of the second step, the estimated clusters are again  frozen as
``regional" clusters.

\underline{Step 3.}
At the last step, all regional clusters are collected to form the
items for the next, third, step. Again items are split into $M_3$
shards and clustered within each shard.
In the example of Figure 1, $M_3=1$ and iteration stops.
In general, iteration continues until the number of items is
sufficiently small to be clustered in a single shard. Importantly, at each step one need to only scan through a small number of items (created by previous
steps) instead of every observation in a large dataset.   

Each step can be implemented in parallel using instances of the same
MCMC algorithm which takes as input a set of (current) items,
generically denoted by $\ytb=\{\ytb_1,\dots,\ytb_B\}$, and outputs
estimated clusters of these items. Those clusters
then define the items for the next step of the
algorithm. Initially, in step 1, $\ytb_i=\yb_i$ are the original data.
Let $\rt_i=|\ytb_i|$, $i=1,\dots,B$, denote the size of each item, in
terms of number of original data that form $\ytb_i$, and let
$\rtb=\{\rt_1,\dots,\rt_B \}$. 
%similar to before, let $\nt_c$ denote the cardinality of
%cluster $\St_c$  (now a cluster of block units).
%And we use  and similarly for
%$\ntb$. \yy Also, we denote $\rhotb$ the new partition of blocks. \jj  
% and $\bm{n}^{-i}=\bm{n}\backslash
% n_i$. %$\bm{n}^{-i}=\{n_1,\dots,n_{s_i-1},n_{s_i}-1,n_{s_i+1},\dots,n_B \}$. 

\paragraph*{Posterior probabilities for clustering sets of observations.}
In each of the $K$ steps,
the MCMC algorithm iterates between (i) updating the cluster membership,
and (ii) updating cluster-specific parameters given the cluster
membership. The key quantity in updating the
cluster membership is the conditional probability 
\begin{eqnarray}
  \label{eqn:cp}
  p(\st_i=c \mid \stib,\ytb,\ntb,\thtsb)\propto
      p(\st_i=c \mid \stib,\ntb)p(\ytb_i \mid \thts_{c})
\end{eqnarray}
for $i=1,\dots,B$ and $c=1,\dots,C^{-i}+1$ where  $\st_i =c$ means that item $i$ is in cluster $c$, i.e.,
all observations in $\ytb_i$ are assigned to cluster $c$.
The definition of the items $\ytb_i$ and the number of
items, $B$, changes across steps. Initially, $\ytb_i$ are the original
data, and $B=n$ is the sample size. In step 2, the items $\ytb_i$ are the
local clusters and $B$ is the total number of local clusters, etc.
Importantly, the probabilities that are evaluated under
\eqref{eqn:cp} and used for clustering in steps 1 through 3 are all
based on the same BNP mixture model for the original observations.

Equation \eqref{eqn:cp} states the implied probabilities for
combining clusters of observations into larger clusters.  
The first factor can be evaluated as
\begin{eqnarray}
  \label{cm}
  p(\st_i=c \mid \stib,\ntb)\propto \frac{p(\rho^{+c}
  \mid \ntb)}{p(\rhot^{-i}\mid \ntb^{-i})}
\end{eqnarray}
where $\rho^{+c} =(\rhot^{-i} \backslash \St_c^{-i}) \cup ( \St_c^{-i} \cup \tilde{i})$ is
the new partition that assigns the $i$th item to cluster $c$ (together with all original data that make up $\ytb_i$). The
partition probabilities on the right-hand side of (\ref{cm}) depend
on $\rtb, \ntb$ and the BNP prior $H$.
For example, using $H=PY(\alpha,d,G_0)$ with
concentration parameter $\alpha$, discount parameter $d$ and baseline
probability measure $G_0$ defines the prior partition,
 \begin{eqnarray}
 \label{pppm}
 p(\rhot  \mid \ntb)\propto (\alpha \mid d)_C\prod_{c=1}^C(1-d)_{\nt_c-1},
 \end{eqnarray}
 % $\nt_c=\sum_{i\in \St_c }n_i$ is the size of the $c$th cluster,
where
$(x)_n=x(x+1)\dots(x+n-1)$ denotes the Pochhammer symbol of a rising
factorial,  and $(x \mid y)_n=x(x+y)\dots(x+(n-1)y)$ denotes the Pochhammer
symbol with increment $y$. Substituting \eqref{pppm} into (\ref{cm})
yields 
 \begin{eqnarray}
  \label{eqn:cm}
  p(\st_i=c \mid \stib,\ntb)\propto\left\{
  \begin{array}{lcl}
    \frac{\Gamma(\nt_c^{-i}+\rt_i-d)}{\Gamma(\nt_c^{-i}-d)}
    & \mbox{if}&c=1,\dots,C^{-i}\\
    \frac{(\alpha+dC^{-i})\Gamma(\rt_i-d)}{\Gamma(1-d)}
    & \mbox{if}&c=C^{-i}+1 \end{array}
  \right.,
 \end{eqnarray}
 where $\nt_c^{-i}$ is the size of the $c$th
 cluster after removing the $i$th  item  $\ytb_i$ (recall that size is
 recorded in original data units). In the special case when
 $\rt_i=|\ytb_i|=1$ for all $i$, equation (\ref{eqn:cm}) 
 reduces to the P\'olya urn representation of the PY process.

The second factor in (\ref{eqn:cp}) is the sampling model
evaluated for $\ytb_i$ given the cluster-specific
parameters, which is straightforward to compute (see below for new
empty clusters). Note that SIGN does not reduce the cost of evaluating the likelihood, however it significantly reduces the number of evaluations.   The only
remaining parameters to be sampled in the MCMC are the
cluster-specific parameters $\theta^\star_j$.
Following Algorithm 8 in \cite{neal2000markov}, a value 
$\theta_{C^{-i}+1}^\star$ for a potential new cluster is
generated from the prior distribution.
In the implementation, the case
when resampling $\st_i$ removes a current cluster, say $S_c$, by
re-assigning the only element of a singleton cluster, needs careful attention.
In that case, the cluster-specific parameter $\thts_{c}$ needs to be kept for
possible later use when a new cluster is considered again.
At the end of each MCMC pass,  we compute a least-squares
estimate of the partition \citep{dahl2006model}. 
Algorithm \ref{alg_MCMC} summarizes the scheme.
% Therefore the resulting conditional probability of cluster membership is given by
% \begin{eqnarray}
% p(s_i \mid \stib,\ytb,\ntb,\thtsb)\propto\left\{\begin{array}{cc} p(s_i \mid \stib,\ntb)p(\ytb_i \mid s_i,\theta_{s_i}^\star)\right.
% \end{eqnarray}
   \begin{algorithm}
   	\caption{MCMC}
   	\label{alg_MCMC}
   	\begin{algorithmic}[1]
   		\Function{MCMC}{$\bm{\ytb}$} ~~~~~~~~~~~~~~~~// $\ytb:=\{\ytb_1,\dots,\ytb_B\}$
   		\State Initialize the partition
   		\For{iter$=1,\dots,N$}~~~~~~~~~~~~//$N$: number of iterations
   		\State Update cluster-specific parameters 
                $\th^\star_j$ 
                \State Update cluster membership,   $\st_i$, 
                   using (\ref{eqn:cp})
		\EndFor
   		\State Compute the estimated partition $\widehat{\rho}=\{S_1,\dots,S_C\}$
   		\State \textbf{Output:} 
                $\ytb^*=\{\ytb_1^*,\dots,\ytb_{C}^*\}$~~~ 
                   //$\ytb_c^*:= \bigcup_{i \in S_c} \yt_i$  % \{y_i|i\in S_c\}$
   		\EndFunction
   	\end{algorithmic}
   \end{algorithm}
The complete SIGN algorithm simply repeatedly distributes the items (i.e., blocked observations) into shards and applies Algorithm \ref{alg_MCMC} to each
shard in parallel. The number $K$ of steps is dynamically
determined by specifying a maximum number $R$ (typically a few
hundred) of items that can be clustered in
one processor. Simulation terminates when the total number of items is
less than $R$. The complete scheme is summarized 
in Algorithm \ref{alg_bnp}. SIGN implements approximate inference in the sense that the observations in the same item will not be split in any of the subsequent steps. We empirically examine the accuracy of the approximate inference compared to the full posterior inference in Section \ref{sec:cs}.
\begin{algorithm}
	\caption{SIGN for BNP clustering}
	\label{alg_bnp}
	\begin{algorithmic}[1]
		\Function{SIGN}{$\bm{y},R$}~~~~~~~~~~~~~~~~~~~~~~~~~~~~~// $\bm{y}:=\{y_1,\dots,y_n\}$
		\State Initialize $\ytb=\bm{y}$
		\While{$B >R$}~~~~~~~~~~~~~~~~~~~~~~~~~~~~~~~~// $B$: number of blocks in $\ytb$
		\State Set $M=\ceil{\frac{B}{R}}$~~~~~~~~~~~~~~~~~~~~~~~~~~~~~~~~// $M$: number of shards
		\State Randomly distribute $\ytb$ into $M$ shards: $\ytb_1,\dots,\ytb_{M}\subseteq \ytb$
		\ParFor{each shard $m=1,\dots,M$}~// parallel for loop
		\State $\ytb_m^*=\mbox{MCMC}(\ytb_m)$~~~~~~~~~~~~~~~~~~~~// call Algorithm 1
		\EndParFor
		\State Set $\ytb=\cup_{m=1}^{M}\ytb_m^*$ and redefine $B$~~~~~~// clusters become blocks for next step
		\EndWhile
		\State \textbf{Output:} $\ytb$
		\EndFunction
	\end{algorithmic}
\end{algorithm}

\section{Clustering and classification with PPMx}
\label{sec:ppmx}
The SIGN algorithm can be applied with a wide range of BNP
mixture models. In this paper, we specifically consider the PPMx model that
allows for simultaneously partitioning of heterogeneous
samples and predicting 
outcomes on the basis of covariates.  

To fix notation, let $z_i\in \{0,1\}$ denote a binary outcome
(reserving notation $y_i$ for a later introduced augmented response).
Let $\xb_i=\{\wb_i,\ub_i\}$ denote a set of continuous covariates
$\wb_i=(w_{i1},\dots,w_{ip})$ and a set of categorical covariates $\ub_i=(u_{i1},\dots,u_{iq})$ for experimental units $i=1,\dots,n$. Let
$\bm{z}=\{z_1,\dots,z_n\}$ and $\bm{x}=\{\xb_1,\dots,\xb_n \}$. A
product partition model (PPM) \citep{hartigan1990partition} assumes
% \begin{eqnarray*}
$p(\rho)\propto \prod_{c=1}^Ch(S_c)$,
% %p(\rho \mid \bm{x})\propto \prod_{c=1}^Ch(S_c)g(\bm{x}_c^\star),
% \end{eqnarray*}
where $h(\cdot)$ is a non-negative cohesion function that quantifies
the tightness of a cluster. For example, the prior
distribution on partitions that is induced under
i.i.d. sampling from a DP-distributed random measure with concentration parameter
$\alpha$ is a PPM with
$h(S_c)=\alpha\times( \mid S_c \mid -1)!$. \cite{muller2011product} define the PPMx as a variation of the 
PPM by introducing prior dependence on covariates by augmenting
the random partition to
\begin{eqnarray}
\label{eqn:ppmx}
p(\rho \mid \bm{x})\propto \prod_{c=1}^Ch(S_c)g(\bm{x}_c^\star),
\end{eqnarray}
with a nonnegative similarity function $g(\cdot)$ indexed by covariates where $\bm{x}_c^\star=\{\xb_i \mid i\in S_c\}$ are the covariates
of observations in the $c$th cluster. The similarity function measures
how similar the covariates are thought to be. A computationally convenient default way to define a similarity
function uses the marginal probability in an
auxiliary probability model $q$ on $x$:
$$
g(\xb_c^\star)=\int\prod_{i\in S_c}q_x(\xb_i \mid \xib_c)q_\xi(\xib_c)d\xib_c.
$$
The important feature here is that the marginal distribution has higher density value for a set of very similar $x_i$ than for a very diverse set.
For continuous covariates, we use an independent normal-normal-gamma auxiliary model,
$q_x(w_{ij} \mid \mu_c,\lambda_c)=\mbox{N}(w_{ij} \mid \mu_c,\lambda_c^{-1})$
and
$q_\xi(\mu_c,\lambda_c)=\mbox{N}(\mu_c \mid \mu_0,(v_0\lambda_c)^{-1})\times
\mbox{Ga}(\lambda_c \mid a_\lambda,b_\lambda).$ For categorical covariates
with $r$ categories, we use a categorical-Dirichlet auxiliary model,
$q_x(u_{ij} \mid \bm{\pi}_c)=\mbox{Cat}(u_{ij} \mid \bm{\pi}_c)$ and
$q_\xi(\bm{\pi}_c)=\mbox{Dir}\left(\bm{\pi}_c\mid
a_\pi,\dots,a_\pi\right)$ with
$\bm{\pi}_c=(\pi_{c1},\dots,\pi_{cr})$. 
The prior $p(\rho \mid \xb)$ introduces the desired
covariate-dependent prior on the clusters $S_c$. Conditional on $\rho$
we introduce cluster-specific parameters $\betab_c$ and complete the model with a probit sampling model,
\begin{eqnarray}
  \label{eqn:like}
  p(\bm{z} \mid \rho,\betab,\xb)=\prod_{c=1}^C\prod_{i \in S_c}p(z_i \mid \xb_i,\betab_c)=\prod_{c=1}^C\prod_{i \in S_c}p_i^{z_i}(1-p_i)^{1-z_i}
\end{eqnarray}
with $p_i=\Phi(\xb_i\betab_c)$ and a centered multivariate normal
prior on $\betab_c\sim \mbox{N}(0,\tau_\beta I)$.

A practical advantage of
the PPMx is its simple implementation. The posterior defined by
models (\ref{eqn:ppmx}) and (\ref{eqn:like}) becomes
$$
    p(\rho,\betab,\xib \mid \bm{z},\bm{x}) 
    \propto\prod_{c=1}^C\left[\left\{\prod_{i\in S_c}p(z_i \mid
      \xb_i,\betab_c)q_x(\xb_i \mid
      \xib_c)\right\}p(\betab_c)q_\xi(\xib_c)h(S_c)\right].
    %% &:=\longunderbrace{\prod_{c=1}^C\prod_{i\in S_c} q_y(\yb_i \mid
    %%   \thetab_c^\star)}{auxiliary likelihood}\times
    %% \longoverbrace{\prod_{c=1}^C q_\theta(\thetab_c^\star)}{parameter
    %%   prior}\times \longunderbrace{q_\rho(\rho)}{partition prior}  
$$
Letting $\yb_i=\{z_i,\xb_i\}$, $\thetab_c^\star=\{\betab_c,\xib_c\}$,
$q_y(y_i \mid \thetab_c^\star)=p(z_i \mid \xb_i,\betab_c)q_x(\xb_i
\mid \xib_c)$,
$q_\theta(\thetab_c^\star)=p(\betab_c)q_\xi(\xib_c)$
and $q_\rho(\rho)=\prod_{c=1}^Ch(S_c)$ one can rewrite the posterior
distribution as
\begin{equation}
  \label{eqn:auxmo}
    p(\rho,\betab,\xib \mid \bm{z},\bm{x}) \propto
   \prod_{c=1}^C\prod_{i\in S_c} q_y(\yb_i \mid \thetab_c^\star)
    \times \prod_{c=1}^C q_\theta(\thetab_c^\star)
    \times q_\rho(\rho).
\end{equation}
That is, posterior inference can proceed as if $\yb_i$ were
sampled from Equation \eqref{eqn:auxmo}. For example, in our
application, we choose $q_\rho(\cdot)$ to be the random partition that
is induced by a PY prior. The PY process
generalizes the DP and is more flexible in modeling the number
of clusters \citep{de2015gibbs}. Posterior inference under
(\ref{eqn:auxmo}) can then be carried out 
using Equation (\ref{eqn:mixt}) (and hence Algorithms \ref{alg_MCMC} and
\ref{alg_bnp}) with $p(y_i \mid \cdot)=q_y(y_i \mid \cdot)$,
$H=PY(\alpha,d,G_0)$ and $G_0=q_\theta(\cdot)$. Note how
\eqref{eqn:auxmo} is identical to the posterior in a 
model with data $\yb_i$, cluster-specific parameters $\thsb_c$ and
prior $q_\rho(\rho)$, allowing for easy posterior simulation.
%We call the implementation of PPMx with SIGN, PPMx\_SIGN.

One of the goals in our later applications is to classify a new
subject, i.e., predict the binary outcome $z_{n+1}$, on the basis
of covariates $\xb_{n+1}$. It is
straightforward to predict $z_{n+1}$ using posterior averaging with respect to partitions, cluster
allocation and model parameters. Let $q(\xb_{n+1} \mid
\bm{x}_c^\star)=g(\bm{x}_c^\star,\xb_{n+1})/g(\bm{x}_c^\star)$. The
posterior predictive distribution is given by 
\begin{equation*}
\label{eqn:class}
\begin{aligned}
  p(z_{n+1} \mid \xb_{n+1},\bm{z},\xb)&\propto
   \int\bigg\{ (\nt_c-d)\sum_{c=1}^{C}p(z_{n+1} \mid \xb_{n+1},\betab_c,s_{n+1}=c)q(\xb_{n+1} \mid \bm{x}_c^\star)\\
&+(\alpha+dC)p(z_{n+1} \mid \xb_{n+1},\betab_{C+1})g(\xb_{n+1})\bigg\}p(\rho \mid \bm{z},\bm{x})d\rho,
\end{aligned}
\end{equation*}
which can be approximated by
\begin{equation*}
\label{eqn:approx}
\begin{aligned}
p(z_{n+1} \mid \xb_{n+1},\bm{z},\bm{x})&\propto \frac{1}{T}\sum_{t=1}^T\bigg\{ (\nt_c^{(t)}-d)\sum_{c=1}^{C^{(t)}}p(z_{n+1} \mid \xb_{n+1},\betab_c^{(t)},s_{n+1}^{(t)}=c)q(\xb_{n+1} \mid \bm{x}_c^\star)\\
&+(\alpha+dC^{(t)})p(z_{n+1} \mid \xb_{n+1},\betab_{C+1}^{(t)})g(\xb_{n+1})\bigg\},
\end{aligned}
\end{equation*}
with superscript $(t)$ indexing $t$th MCMC samples, $t=1,\dots,T$,
and $\betab_{C+1}^{(t)}$ is drawn from its prior. 

\section{Simulation}
\label{sec:cs}
 % In this section, we
%% % test the proposed SIGN algorithm in four case studies. We
%% consider two simulated data
%% % in the first two studies and
%% applications to EHR data and Bank deposit data in the third and fourth
%% studies. For simulated data, we
We consider simulations with relatively small datasets
with $n=800$, $p=5$ and $q=5$, such that we can compare with a standard MCMC implementation of PPMx. The scalability is explored later in  two case studies.
% Bank telemarketing studies.
 We report frequentist summaries based on 50 repetitions. Throughout
this section, we set the hyperparameters at $\alpha=1,
d=0.5,\tau_\beta=1,\mu_0=0,v_0=a_\lambda=b_\lambda=0.01,a_\pi=1/r$.
MCMC is run for 10,000 iterations at each step. We discard the first 50\% of MCMC
samples as burn-in and thin the chain by keeping every 5th sample.

\subsection{Simulation I: cluster-specific probit regression}
\label{sec:sd1}
\begin{figure}[h]
	\centering
	\subfigure{\includegraphics[width=.49\textwidth]{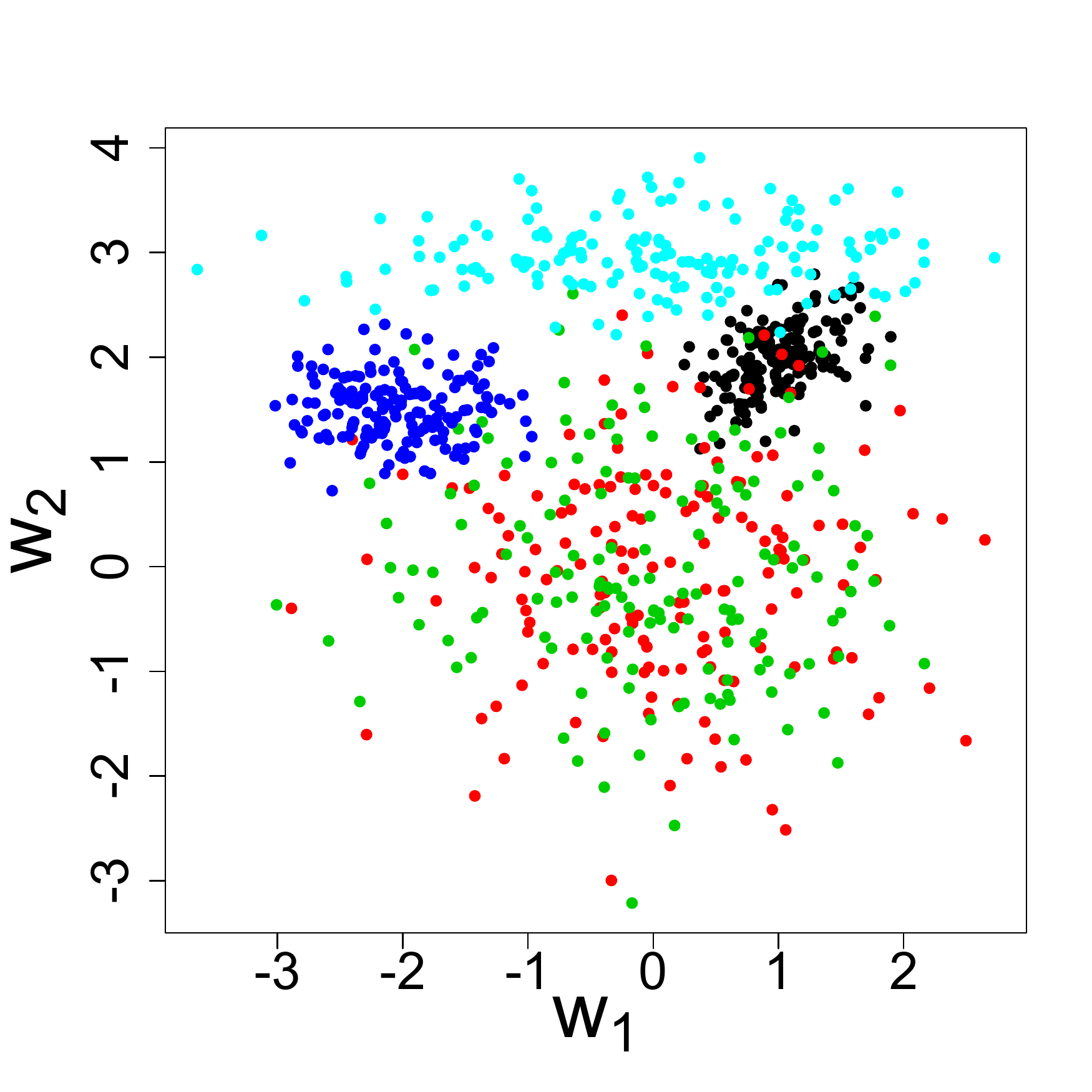}}
	\subfigure{\includegraphics[width=.49\textwidth]{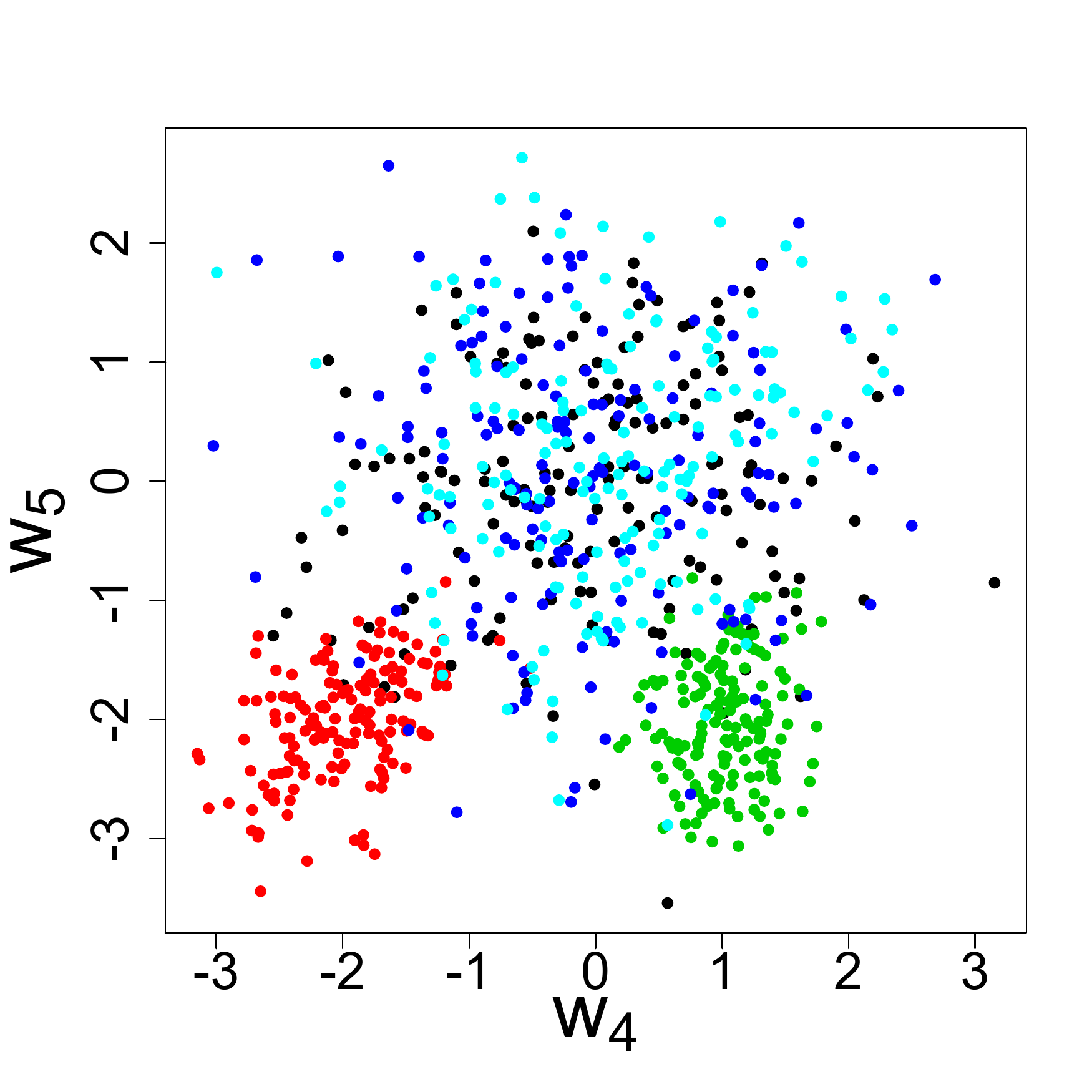}}
	\caption{A typical view of data from simulation I.}
	\label{w1234}
\end{figure}

We consider
 a scenario where the simulation truth includes underlying
clusters. 
In particular, we assume
 a simulation truth with $C_0=5$ clusters, and all clusters having 
the same size. 
Discrete
covariates $\ub_i$ are  generated as
$u_{ij}\sim Cat(1/3,1/3,1/3)$, independently,
$j=1,\dots,q$. Continuous 
covariates $\wb_{i}$ are generated from $N_p(\mub_c,\Sigmab_c)$ given
$s_i=c$, where $\mub_1=(-2,1.5,0,0,0)^T$, $\mub_2=(0,4,0,0,0)^T$,
$\mub_3=(0,0,0,1,-2)^T$, 
$\mub_4=(1,2,0,0,0)^T$, $\mub_5=(0,0,0,-2,-2)^T$,
$\Sigmab_1=diag(0.25,0.05^2,1,1,1)$,
$\Sigmab_2=diag(1.25^2,0.05^2,1,1,1)$,
$\Sigmab_3=diag(1,1,1,0.05^2,0.25)$, 
\[\Sigmab_4=blkdiag\left(\left[\begin{array}{cc}0.1&0.05\\0.05&0.1\end{array} \right],I_3\right)\mbox{~and~} \Sigmab_5=blkdiag\left(I_3,\left[\begin{array}{cc}0.25&0.125\\0.125&0.25\end{array} \right]\right).\]
In words, clusters 1, 2, and 4 are characterized by a shift in
the distribution for the first two continuous covariates $w_{i1}$
and $w_{i2}$ with different correlation 
structures whereas clusters 3 and 5 are characterized by a shift
in the third and fourth continuous covariates $w_{i4}$ and
$w_{i5}$. And $w_{i3}$ plays the role of a ``noisy'' covariate with the same
distribution across all clusters.
Covariates that do not define the clusters (such as
$w_{i3},w_{i4},w_{i5}$ in clusters 1, 2 and 4) are independently
sampled from standard normal distributions. A typical view of the data
is shown in Figure \ref{w1234} where we plot $w_{i1}$ v.s. $w_{i2}$
and $w_{i3}$ v.s. $w_{i4}$. The binary response $z_i$ is generated
from  a  cluster-specific probit regression, $z_i\sim Bernoulli(p_i)$ with 
\begin{eqnarray*}
	\Phi^{-1}(p_i)=\left\{\begin{array}{lcl}-1-w_{i5}&\mbox{if}&s_i=1\\-1+2w_{i3}&\mbox{if}&s_i=2\\-1+w_{i4}&\mbox{if}&s_i=3\\-1+1.5w_{i1}-I(u_{i1}=2)+I(u_{i1}=3)&\mbox{if}&s_i=4\\-1-1.5w_{i1}-I(u_{i2}=2)+I(u_{i3}=3)&\mbox{if}&s_i=5\\\end{array}\right.
\end{eqnarray*}
\indent
We carry out inference under the PPMx model using the default
similarity functions (simply PPMx
hereafter) and use a SIGN implementation with $K=2$ steps. In the first step
of SIGN, the training samples are randomly  split into  $M_1=4$
shards with each shard processing 200 samples. For comparison
we also carry out inference using k-means 
\citep{hartigan1979algorithm} for the continuous covariates (which
define the clusters) with $k=5$ (the true number of clusters) and 20
random starting points. PPMx is always able to correctly identify the
number of clusters with 5\% average misclassification rate (with respect to
cluster assignment). SIGN selects 
the correct number of clusters in 48 (out of 50) simulations for which
the average misclustering rate is 15\%. In contrast, with k-means we find
a misclassification rate of 51\%. \\ 
\indent To assess the out-of-sample predictive performance, 
 that is, prediction of $z_{n+1}$, 
we compute
the area under the ROC curve (AUC) based on 50
independent test samples generated from the same simulation
truth as the 
training data. In addition to the comparison with PPMx, we also
benchmark SIGN against four more alternative classifiers: sparse LR with
lasso (R package {\tt "glmnet"}), SVM ({\tt "e1071"}), RF ({\tt "randomForest"}) and BART ({\tt "BayesTree"}). For
SVM, we transform the discrete covariates using dummy variables, fit
with linear, cubic and Gaussian radial basis and report the best
performance of the three. We grow 50,000 trees for RF and 200 trees
for BART. For a fair comparison, we run BART using the same MCMC configuration
as ours (i.e. 10,000 iteration, 50\% burn-in and save every 5th
sample).  The results are reported in the first column of
Table \ref{tab:auc} where we find SIGN and PPMx have almost the same
AUC's and both compare favorably with the competing classifiers.  
%DT ({\tt"C50"})
%DT is implemented using Quinlan's C5.0 algorithm \citep{quinlan2014c4}.
%and CART ({\tt "party"})

\subsection{Simulation II: non-linear probit regression}
\label{sec:sd2}
The favorable results for SIGN and PPMx in simulation I
may be partially due to the  chosen simulation truth.  
For an alternative comparison, in this
example we use a simulation truth different from the PPMx
model. Particularly, we assume  a simulation truth without an underlying clustering structure, and we generate the binary response by a
nonlinear probit regression, $z_i\sim Bernoulli(p_i)$ with 
\begin{multline}
\Phi^{-1}(p_i)=-1+w_{i1}^2-w_{i2}^2+\sin(w_{i3}w_{i4})+
I(u_{i1}=2)-I(u_{i1}=3)-I(u_{i2}=2)+I(u_{i2}=3).
\nonumber
\end{multline}
The AUC summaries for the classification are shown in the second column of Table
\ref{tab:auc}. SIGN, PPMx and RF have the same AUC, $AUC=0.84$, 
which is slightly lower than the AUC of BART, $AUC=0.87$. 
LR and SVM do not perform well in
both simulations possibly due to the parametric (linear or cubic)
decision boundary in LR and the use of SVM with linear and cubic bases, and the
difficulty in tuning the model parameters in SVM with radial bases.

\section{Case studies}
\label{sec:cs2}
\subsection{Electronic health records data: detecting diabetes}
\label{sec:app}
EHR data provide great opportunities as well as challenges for
data-driven approaches in  early  disease detection.  Large sample sizes
allow more efficient statistical 
inference but at the same time impose computational challenges, especially
for flexible  but computation-intensive BNP models.

We consider EHR data for $n=85,021$ individuals in China. The dataset
is based on a physical examination of residents in some districts of a
major city in China 
conducted in 2016. We use the data to develop a model for chronic disease
prediction, specifically for diabetes. We extract data on diabetes
from the items ``medical history'' and ``other current 
diseases'' in the physical examination form. If either of the two
items of a subject contain diabetes, that subject is
considered as having diabetes. We denote the diabetes status by $z_i$ (1:
diabetic and 0: normal) for subjects 
$i=1,\dots,n$. Blood samples were drawn from each subject and sent to a
laboratory for subsequent tests. We consider test results that are
thought to be relevant to diabetes. These include white blood cell
count (WBC), red blood cell count (RBC), hemoglobin (HGB), platelets
(PLT), fasting blood glucose (FBG), low density lipoproteins (LDL),
total cholesterol (TC), triglycerides (Trig), triketopurine (Trik),
high density lipoproteins (HDL), serum creatinine (SCr), serum
glutamic oxaloacetic transaminase (SGOT) and total bilirubin (TB). We
also include 5 additional covariates: gender, height, weight,
blood pressure and waist. Our goal is two-fold: (1) predicting
diabetes; and
(2) clustering a heterogeneous population into homogeneous
subpopulations.

\paragraph*{GAN preprocessing.}
To comply with Chinese policy, we report inference for data generated
by a Generative Adversarial Network (GAN,
\citealt{goodfellow2014generative}), which replicates  the
distribution underlying the raw data, 
GAN is a machine learning algorithm which
simultaneously trains a generative model and a discriminative model on
a training dataset (in our case, the raw EHR dataset). The generative
model simulates the training data distribution in order to “trick” the
discriminative model. Meanwhile, the discriminative model learns to
optimally distinguish between data and simulations. During training,
the generative model uses gradient information from the discriminative
model to produce better simulations. After training, the generative
model can be used to generate an arbitrary number of simulations which
are similar in distribution to the original dataset. In our case, we
generate a simulated dataset of the same size as the raw EHR dataset.

For this application, we train on a dataset where columns of
continuous variables are standardized, and corresponding output are
then re-scaled at simulation time. To accommodate binary variables, we
allow the GAN to simulate continuous values, and round corresponding
outputs to 0 or 1. We use the architecture of MMD-GAN
\citep{li2017mmd}, which uses the maximum mean discrepancy
\citep[MMD,][]{gretton2012kernel}, a distributional distance, to
compare real data and simulations. Our implementation uses encoder and
decoder networks each containing three layers of 100 nodes, connected
by a bottleneck layer of 64 nodes, and with exponential linear unit
activations. In our optimization, we use RMSProp with a learning rate
of 0.001, and we weight the MMD in our discriminator loss function by
0.1. 

\begin{figure}[h]
	\centering
	\subfigure[Marginal distributions]{\includegraphics[width=.6\textwidth]{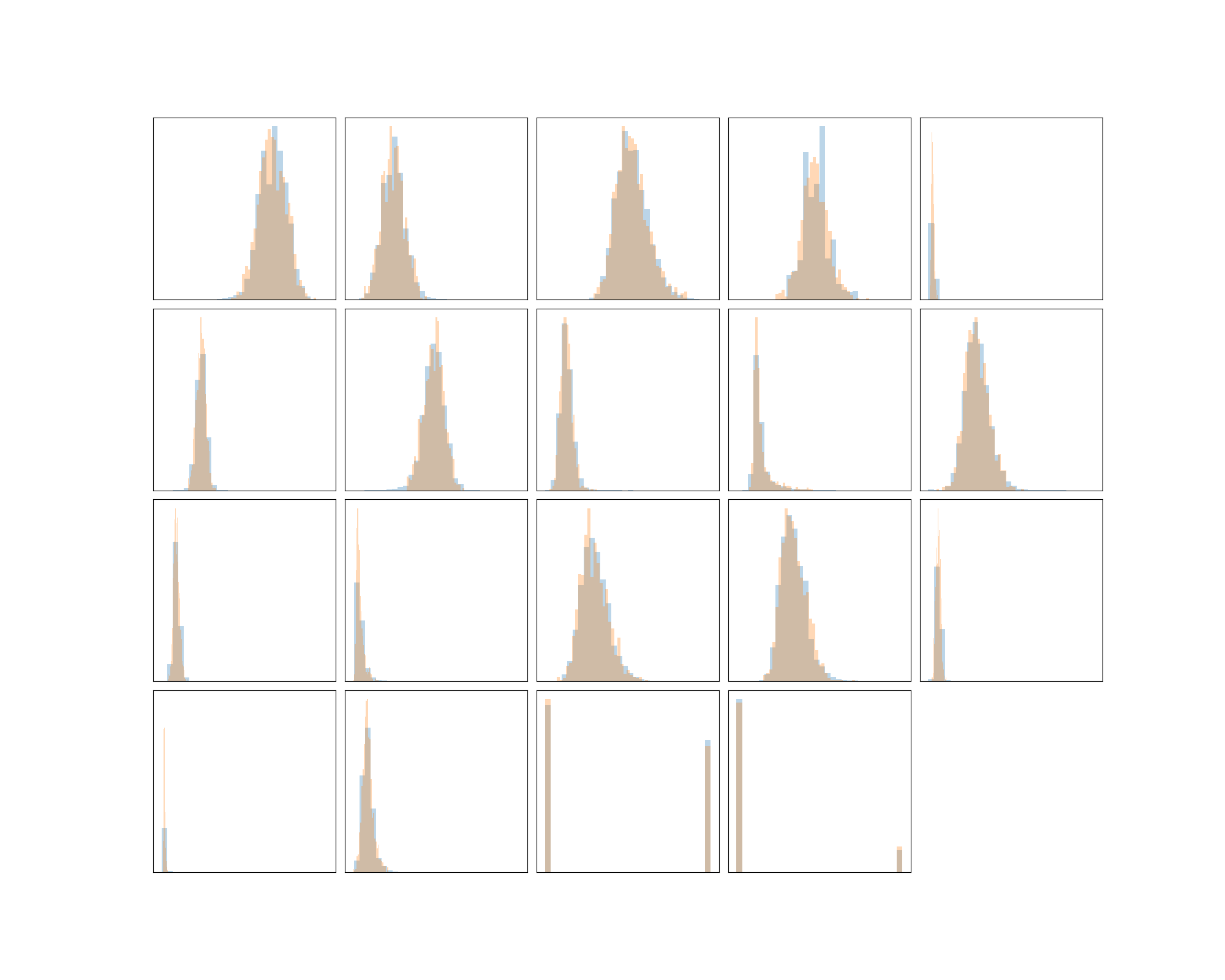}}
	\subfigure[Pairwise correlations]{\includegraphics[width=.39\textwidth]{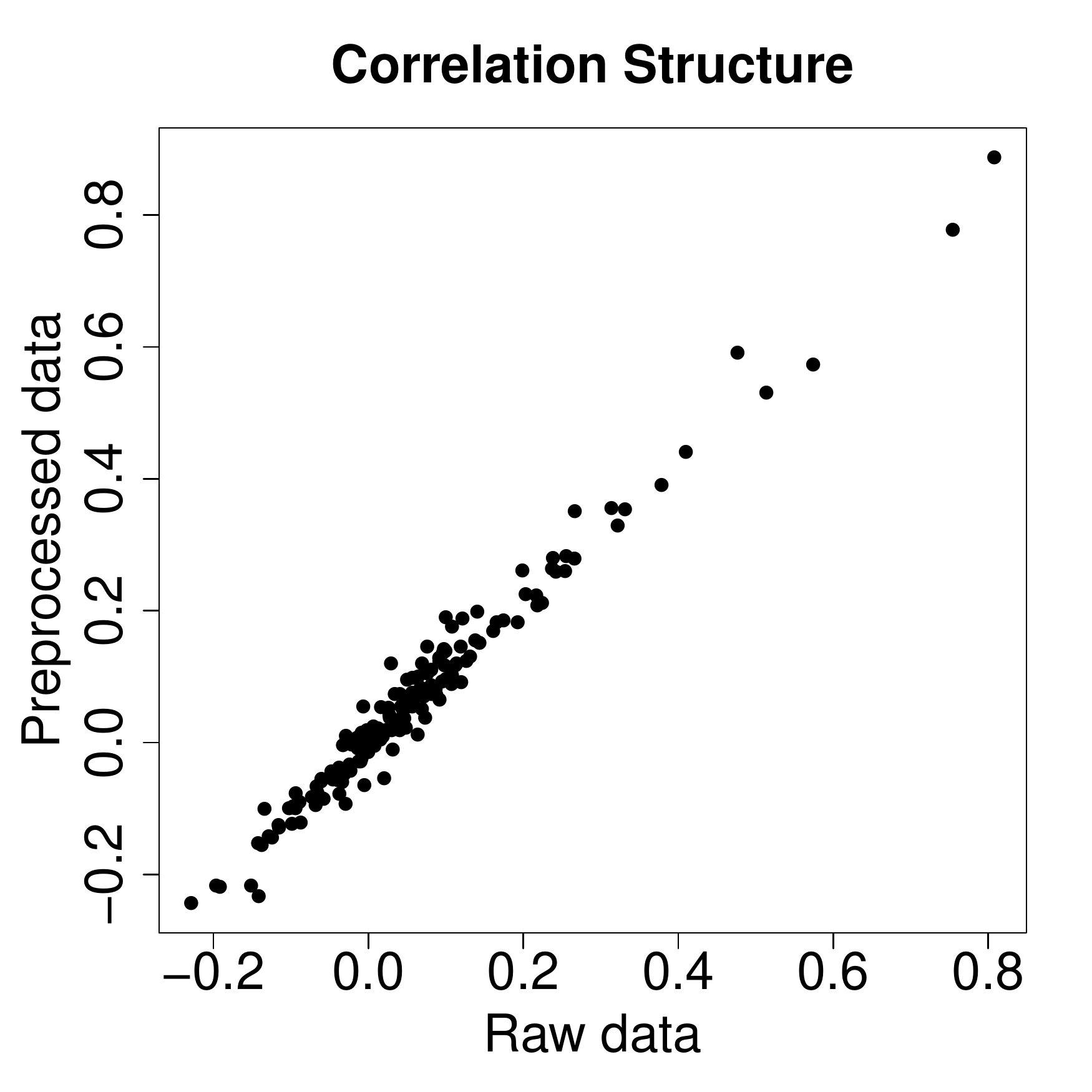}}
	\caption{GAN-preprocessed EHR data versus raw EHR data. (a)
          Marginal distribution of each variable. 
        	 For each variable, the two overlaid histograms show the agreement between the preprocessed and the raw data.
          The variable names
          and ranges are deliberately not shown. (b) Correlation of
          each pair of variables. Each dot represents the Pearson correlation coefficients of one pair of variables in the raw EHR data (x-axis) versus in the GAN-preprocessed EHR data (y-axis). In total, we have ${19\choose 2}$ pairs/dots.} 
	\label{mar_cor}
\end{figure}

Our model reaches a stable point, where both marginal distributions and pairwise
correlations agree with the raw data (see Figure \ref{mar_cor}). Moreover, the classifiers we consider
have similar prediction performance on the two datasets. Therefore, we
only report the results based on the replicated EHR data
(referred to as EHR data hereafter).  To the extent to which the
preprocessed data set retains all 
information and structure of the original data, any inference other
than subject-specific summaries remains practically unchanged. See the Appendix for more details.

\paragraph*{Results.}
We randomly sample  84,750 subjects as training data and use the
remaining 271 subjects as test data to evaluate out-of-sample
classification performance. We implement inference under
the PPMx model using the proposed SIGN algorithm.
% the standard implementation is not used here because it is expected to
% take more than a week. 
In the implementation, we use 250 compute cores (equivalent to 11
compute nodes with 24 cores per node) at the Texas 
Advanced Computing Center (TACC, \url{http://www.tacc.utexas.edu}) for
computation.   In the first step, the training samples are randomly
split into those $M_1=250$ compute cores/shards with 339 samples on
each shard. Collectively, we obtain 1351 local clusters. In the  
second step, the 1351 local clusters are distributed to $M_2=5$
shards with each shard taking about 270 items. The local
clusters are grouped into 25 regional clusters. Iteration stops there since $25$ items need not be further
split, i.e., $K=3$.
% small enough to be efficiently processed on one shard, 
In a final step, the  25 regional clusters are merged
to 5 global clusters 
with sizes 26892, 26453, 18778, 11474 and 1153.    
%\begin{figure}[h]
%	\centering
%	\includegraphics[width=\textwidth]{figs/noc2}
%	\caption{EHR. Number of clusters across shards for Step 1 (solid, left y-axis) and Step 2 (dashed, right y-axis).}
%	\label{noc}
%\end{figure}

The AUC summaries based on the test dataset are
provided in Table \ref{tab:auc}. SIGN reports the highest AUC (0.880)
followed by RF and BART. As expected, the most important covariate for
predicting diabetes is FBG. Regressing on FBG alone achieves $AUC=0.829$.
In terms of computation time, SIGN, BART, RF 
and SVM take 0.9, 18.7, 3.5 and 2.5 hours with 2.6 GHz Xeon E5-2690 v3
CPU, respectively, whereas LR is several magnitude faster at the
price of accuracy. We do not implement PPMx with standard
MCMC, as this is not feasible with the large sample size.
\begin{table}
	\caption{Performance of the methods used for Simulations I
		and II, and the two case studies. 
                 The table reports AUC for inference under 
                SIGN, (standard implementation of) PPMx, BART, RF, LR
                and SVM. 
                 Numerical errors (as standard deviations over
                repeat simulation) are 
                given within the parentheses.
             }
	\centering
	\begin{tabular}{ccccc}
		\\\hline\hline
		&Simulation I & Simulation II 
		&EHR&Bank\\\hline
		SIGN&0.808 (0.067) & 0.838 (0.067) &0.880&0.825\\
		PPMx &0.824 (0.060)&0.841 (0.063)&-&-\\
		%CART & 0.683 (0.073) & 0.805 (0.067) &0.845 &0.798\\
		BART &0.755 (0.062)&0.866 (0.050)&0.867&0.792\\
		RF &0.793 (0.059)&0.838 (0.067)&0.869&0.786\\
		%DT &0.714 (0.073)&0.829 (0.067)&0.823&0.788\\
		LR &0.600 (0.091)&0.524 (0.073)&0.856&0.781\\
		SVM &0.622 (0.077)&0.585 (0.077)&0.856&0.761\\\hline
	\end{tabular}
	\label{tab:auc}
\end{table} 
%\begin{table}
%	\caption{Summaries of the performance of the methods used for Simulations I
%          and II, and the two case studies. AUCs of SIGN, (standard implementation of) PPMx, CART, BART, RF, DT, LR and SVM. The standard deviations are given within the parentheses.}
%	\centering
%	\begin{tabular}{ccccc}
%			 \\\hline\hline
%			 &Simulation I & Simulation II 
%       &EHR&Bank\\\hline
%			SIGN&0.808 (0.067) & 0.838 (0.067) &0.848&0.825\\
%			PPMx &0.824 (0.060)&0.841 (0.063)&-&-\\
%			%CART & 0.683 (0.073) & 0.805 (0.067) &0.845 &0.798\\
%			BART &0.755 (0.062)&0.866 (0.050)&0.836&0.792\\
%			RF &0.793 (0.059)&0.838 (0.067)&0.833&0.786\\
%			DT &0.714 (0.073)&0.829 (0.067)&0.823&0.788\\
%			LR &0.600 (0.091)&0.524 (0.073)&0.821&0.781\\
%			SVM &0.622 (0.077)&0.585 (0.077)&0.810&0.761\\\hline
%	\end{tabular}
%	\label{tab:auc}
%\end{table}
The  good performance of SIGN may be explained by its ability
to explicitly 
accommodate the heterogeneous nature of the subject population and
allow for cluster-specific probit models in each subpopulation while
leveraging model averaging to classify new subjects. For example, the
estimated intercept is -1.5 for cluster 2 and -0.95 for cluster 4. The
coefficient of the important covariate FBG also exhibits
heterogeneity, 0.97 for cluster 3 and 0.76 for cluster 4. 

%As expected, the most important covariate for predicting diabetes is FBG. Regressing on FBG alone achieves 0.829 AUC. Although classification rule based on FBG is similar for the entire population, other covariates exhibit heterogeneity across clusters. For example, WBC seems to provide little classification power for the entire dataset (left panel of Figure \ref{x5}), however, in cluster 6 (right panel), we observe a positive association between WBC and diabetes. In fact, WBC has been identified as an independent risk factors for diabetes (especially type II, \citealt{vozarova2002high,twig2013white}).  Similarly, each of clusters 1, 2 and 6 has a distinct pattern in classifying diabetic subjects using Trig (Figure \ref{x12}), another well-known risk factors for diabetes \citep{dotevall2004increased,tirosh2008changes,skretteberg2013triglycerides,lee2014predicting}. 
%\begin{figure}[h]
%	\centering
%	\includegraphics[width=\textwidth]{figs/x9}
%	\caption{EHR. Empirical density of fasting blood glucose for diabetes ($z=1$) vs normal ($z=0$). }
%	\label{x9}
%\end{figure} 
%\begin{figure}[h]
%	\centering
%	\includegraphics[width=\textwidth]{figs/x5}
%	\caption{EHR. Empirical density of white blood cell count for diabetes ($z=1$) vs normal ($z=0$). Left panel: full data. Right panel: cluster 6. }
%	\label{x5}
%\end{figure}
%\begin{figure}[h]
%	\centering
%	\includegraphics[width=\textwidth]{figs/x12}
%	\caption{EHR. Empirical density of triglygerides for diabetes ($z=1$) vs normal ($z=0$). Left, middle and right panels: clusters 1, 2 and 6.}
%	\label{x12}
%\end{figure}
\subsection{Predicting the success of telemarketing} 
Direct marketing is a form of advertising where the salesperson
directly communicates with the customers to promote business. In 2011,
marketers are estimated to have spent \$163 billion on direct
marketing which accounted for 52.1\% of total US advertising
expenditures in that year \citep{direct2012power}. A common direct
marketing practice is  by phone, known as telemarketing. 
In this study, we focus on predicting the success of telemarketing in
selling long-term bank deposits.

\indent We analyze a dataset collected from a Portuguese retail
bank \citep{moro2014data} with $n=41,188$ records. The outcome of
interest is whether the customer  eventually  subscribed a
long-term deposit: $z_i=1$  if yes, and $z_i=0$ otherwise,
$i=1,\dots,n$. Associated with 
each record/customer are 20 covariates which are listed in Table
\ref{tab:cov}. We follow \cite{moro2014data} and remove the covariate
``last contact duration", since the duration is unknown before a call
is performed and therefore can not be used to predict the outcome of the next customer. 
% the outcome $z_i$ is obviously known after the end of
% the call. In other words, to predict the outcome of a future customer,
% the ``last contact duration" cannot be observed before the
% outcome. 
After removing records that are inconsistent with the
data description, the resulting dataset contains 37,078 records. We
randomly sample $n=36,750$ as training data and use the remaining 328 for
testing purpose. Similarly to the analysis in Section \ref{sec:app},
we apply PPMx using SIGN with $K=3$ steps. In the first step, we
randomly split the training data into $M_1=150$ shards (distributed on
7 compute nodes) with each shard taking 245 samples. We find 1,474
local clusters in the first step. Then the 1,474 local clusters are
split to $M_2=5$ shards with each shard processing about 295 blocks of
customers. In this step, the local clusters are merged into 64
regional clusters. Finally, the 64 regional clusters are grouped into 14
global clusters with cluster sizes 7,687, 6,042, 5,725, 5,130, 3,950, 2,815,
2,101, 1,484, 975, 689, 56, 48, 26 and 22.\\ 
\indent The classification performance evaluated on the testing dataset is reported in the last column of Table \ref{tab:auc} for SIGN, BART, RF, LR and SVM. We find SIGN outperforms all other methods with $\mbox{AUC}=0.825$. The second best algorithm is BART with $\mbox{AUC}=0.792$.
\begin{table}
	\caption{20 Covariates in the long-term deposit data. For
          categorical covariates, the number within the parentheses
          indicates the number of categories.} 
	\centering
	\begin{tabular}{ll}
		Covariate name & Type\\\hline
		Type of job&Categorical (12)\\
		Marital status&Categorical (4)\\
		Education&Categorical (8)\\
		Default or not&Categorical (3)\\
		Housing loan or not&Categorical (3)\\
		Contact communication type& Categorical (2)\\
		Last contact month of year&Categorical (12)\\
		Last contact day of the week&Categorical (5)\\
		Outcome of the previous campaign& Categorical (3)\\
		Age&Continuous\\
		Last contact duration&Continuous\\
		Number of contacts & Continuous\\
		Number of days from a previous campaign&Continuous\\
		Number of contacts before this campaign&Continuous\\
		Employment variation rate&Continuous\\
		Consumer price index&Continuous\\
		Consumer confidence index&Continuous\\
		Euribor 3 month rate&Continuous\\
		Number of employees& Continuous
		\end{tabular}
	\label{tab:cov}
\end{table}
%\begin{figure}[h]
%	\centering
%	\includegraphics[width=\textwidth]{figs/noc3}
%	\caption{Telemarketing. Number of clusters across shards for Step 1 (solid, left y-axis) and Step 2 (dashed, right y-axis).}
%	\label{noc3}
%\end{figure}
\section{Discussion}
\label{sec:disc}
We have introduced SIGN as a scalable algorithm for inference
on clustering under BNP mixture models.
SIGN can be thought of as a parallelizable extension of Neal's algorithm
8 which is applicable to both conjugate and non-conjugate models. We use SIGN to implement inference under a PPMx model for a Chinese EHR dataset with 85,021 individuals and a bank telemarketing dataset with
37,078 customers. We find  good  classification performance
compared with the state-of-the-art competing methods. For the EHR
study, we  find five meaningful clusters in the study population. 
% which partially explains the superiority of PPMx combined with
% SIGN. 
We anticipate that this study will continue to collect many more
subjects over the coming years. The use of algorithms that are
scalable to millions of observations in terms of both computing time
and memory is therefore imperative. The computing time for the
proposed algorithm remains practicable with increasing sample size as
long as enough computing resources are available. For example, with
1,000,000 observations, we roughly need to run SIGN for about 1 hour
on 2,000 cores or equivalently around 80 compute nodes.  This is
feasible on many high performance computing centers such as TACC.  And
memory is a lesser issue because if needed one can use one large-memory
compute node (192GB on TACC) in the last step where we have to access
the entire dataset.

In this paper, we only consider ``large $n$, small $p$" problems.
The two motivating applications include only $p=18$ and $p=19$
covariates. Extension to ``large $n$,
large $p$" problems is of high methodological and practical interest
for other problems.  Another limitation of inference for the PPMx
model with the current SIGN implementation is the need to access the
entire dataset in the last step, which becomes computationally
prohibitive for big $n$ or $p$. One possible strategy is to
replace the cluster-specific probit model by a simpler
cluster-specific Bernoulli model for the binary response.
The desired dependence between response and covariate is
introduced marginally, after marginalizing with respect to the
partition. Under this construction the algorithm depends on the data
only through low dimensional summary statistics and could handle
arbitrarily large data.
A similar strategy was explored in \cite{zuanetti2018Bayesian}. 
However, introducing the dependence between response and covariates through
the partition only, we find less favorable classification performance
than in the current implementations (results not shown).
\appendix
\section*{Acknowledgment}
Yang Ni, Peter M\"uller and Yuan Ji's research were partially supported by NIH/NCI grant a R01 CA 132897. Maurice Diesendruck and Sinead Williamson were partially supported by NSF IIS-1447721. The authors acknowledge the TACC at The University of Texas at Austin for providing high performance computing resources that have contributed to the research results reported within this paper. 
\section*{Appendix: GAN preprocessing details}
To evaluate the privacy of the simulated set, we measure two types of risk: presence disclosure and attribute disclosure \citep{choi2017generating}. Presence disclosure is the ability to determine whether a candidate point was included in the training dataset. Attribute disclosure is the ability to predict other attributes of a candidate point, given partial information about that point. For both settings, we choose three sets of equal size -- 5\% of the training data, a heldout set for testing, and a heldout set for baseline comparison -- then estimate the sensitivity and precision of classification schemes. 

For presence disclosure, we sample a candidate from the union of training and testing sets, and classify whether the candidate was included in the training set based on the presence of an $\epsilon$-close neighbor in the simulated set. For large $\epsilon$, the notion of $\epsilon$-closeness is not informative, since many points are returned as neighbors, and precision scores hover around 50\% -- no better than random guessing. For small $\epsilon$, few points are returned as neighbors, and neighbors are more likely to be correctly guessed, since the requirement is for a neighbor to be nearly identical to the candidate point. To reflect the optimal privacy standard, we report scores using the largest $\epsilon$ for which precision exceeds 55\%. This yields the largest sensitivity under non-trivial risk, where a higher sensitivity indicates greater ability to identify a participant. At $\epsilon = 9.5$, the sensitivity of this classification is 0.0005, indicating that compromised training points would be identifiable only 0.05\% of the time.

For attribute disclosure, we sample as above, and classify whether unknown features of a candidate point can be correctly estimated to within 5\% of the true value, by averaging each feature over the candidate's five nearest neighbors in the simulated set. We report values for the case in which half of the candidate's features are known, and the other half are imputed, and note that performance did not change significantly when the percentage of known values differed. The sensitivity and precision scores of this classification are 0.31 and 0.72, respectively, indicating that unknown features would be correctly guessed 31\% of the time, and features claiming to be within 5\% of the true value would in fact be 72\% of the time.

We note that privacy and accuracy goals are inherently opposed. An increase in privacy corresponds to a simulated set with less information about individual data points, and vice versa. As a general guideline, we aim to satisfy privacy requirements while preserving as much as possible the utility of the simulations. In the specific case of attribute risk, we recognize that scores depend on the correlation structure of the data, where highly correlated features are more susceptible to attribute disclosure. As a baseline, we compared attribute risk scores of simulations to those of the final heldout set, and found that both were approximately 30\% and 70\%, respectively.
\bibliographystyle{apalike}
\bibliography{DC_ref}
\end{document}